\documentclass[aps,prd,twocolumn,superscriptaddress]{revtex4-2}

\usepackage{graphicx}
\usepackage{amsmath}
\usepackage{amssymb}
\usepackage{xcolor}
\usepackage{extpfeil}
\usepackage[colorlinks,citecolor=blue,linkcolor=blue,urlcolor=blue,anchorcolor=blue]{hyperref}
\allowdisplaybreaks
\newcommand\dx{{\rm d}}
\newcommand\p{\partial}

\newcommand\AJ{Astron. J.}
\newcommand\ApJ{Astrophys. J.}
\newcommand\ApJL{Astrophys. J. Lett.}

\newcommand\ARAA{Annu. Rev. Astron. Astrophys.}
\newcommand\AstronAstrophys{Astron. Astrophys.}
\newcommand\AstropartPhys{Astropart. Phys.}

\newcommand\ChinPhys{Chin. Phys.}

\newcommand\GravCosmol{Grav. Cosmol.}

\newcommand\JCAP{J. Cosmol. Astropart. Phys.}
\newcommand\JETPLett{JETP Lett.}

\newcommand\JLTP{J. Low Temp. Phys.}
\newcommand\LRR{Living Rev. Relativity}
\newcommand\MNRAS{Mon. Not. R. Astron. Soc.}

\newcommand\MPLA{Mod. Phys. Lett. A}

\newcommand\NSR{Natl. Sci. Rev.}
\newcommand\PDU{Phys. Dark Universe}
\newcommand\PhysRep{Phys. Rep.}

\newcommand\PLB{Phys. Lett. B}
\newcommand\PRD{Phys. Rev. D}
\newcommand\PRL{Phys. Rev. Lett.}

\newcommand\ProgTheorPhys{Prog. Theor. Phys.}

\newcommand\RMP{Rev. Mod. Phys.}

\newcommand\SovPhysJETP{Sov. Phys. JETP}

\begin{document}

\title{Cosmological consequences of a scalar field with oscillating equation of state. III.
   Unifying inflation with dark energy and small tensor-to-scalar ratio}
\author{S. X. Tian}
\email[]{tshuxun@bnu.edu.cn}
\affiliation{Department of Astronomy, Beijing Normal University, 100875 Beijing, China}
\author{Zong-Hong Zhu}
\email[]{zhuzh@bnu.edu.cn}
\affiliation{Department of Astronomy, Beijing Normal University, 100875 Beijing, China}
\affiliation{School of Physics and Technology, Wuhan University, 430072 Wuhan, China}
\date{\today}
\begin{abstract}
  We investigate the inflationary consequences of the oscillating dark energy model proposed by Ti\'an [\href{https://doi.org/10.1103/PhysRevD.101.063531}{Phys. Rev. D {\bf 101}, 063531 (2020)}], which aims to solve the cosmological coincidence problem with multi-accelerating Universe (MAU). We point out that the inflationary dynamics belong to slow-roll inflation. The spectral index of scalar perturbations and the tensor-to-scalar ratio $r$ are shown to be consistent with current \textit{Planck} measurements. Especially, this model predicts $r\sim10^{-7}$, which is far below the observation limits. This result motivates us to explore the smallness of $r$ in the general MAU. We propose a quintessential generalization of the original model and prove $r<0.01$ in general. The null detection to date of primordial gravitational waves provides a circumstantial evidence for the MAU. After the end of inflation, the scalar field rolls toward infinity instead of a local minimum, and meanwhile its equation of state is oscillating with an average value larger than $1/3$. In this framework, we show that gravitational particle creation at the end of inflation is capable of reheating the Universe.
\end{abstract}
\maketitle

\section{Introduction}\label{sec:01}
The present standard cosmological model is known as the Lambda Cold Dark Matter ($\Lambda$CDM) model, in which the cosmological constant $\Lambda$ is generally referred to as the simplest explanation of the late-time cosmic acceleration \cite{Frieman2008.ARAA.46.385,Bartelmann2010.RMP.82.331}. However, simpleness is not an indicator of naturalness, and the latter seems to be more important for a physical theory~\footnote{Simpleness represents mathematical complexity and naturalness represents the probability of choosing by nature.}. The $\Lambda$CDM model is unnatural due to an incredible coincidence: matter dominated the $\Lambda$ by many orders of magnitude through most of the cosmic history, but they are comparable at today \cite{Carroll2001.LRR.4.1}. Dynamical dark energy provides an alternative explanation to the cosmic acceleration, some of which can alleviate the coincidence problem with the tracker property \cite{Peebles1988.ApJL.325.L17,Ratra1988.PRD.37.3406,Steinhardt1999.PRD.59.123504,Zlatev1999.PRL.82.896}. The core argument is that the tracker property allows the initial conditions of dark energy to vary within a wide range without affecting the late-time evolution. However, the problem is not solved as the relative dark energy density is still much less than $1$ through most of the history in the tracker models (see Fig. 6 in \cite{Steinhardt1999.PRD.59.123504} for an example).

Just releasing the initial conditions of dark energy cannot solve the coincidence problem. Intuitively, what we need may be a multi-accelerating Universe (MAU), in which the normal matters and dark energy alternately dominated each other across the whole cosmic history. The present day belongs to one of the transitions, and thus the coincidence problem disappears. This scenario can be realized by oscillating dark energy, which was first studied in \cite{Dodelson2000.PRL.85.5276} with a sinusoidal-modulated exponential potential $V(\phi)\propto\exp(-\lambda\phi)[1+A\sin(B\phi)]$. Follow-up works mainly focus on exploring other forms of the oscillating dark energy, e.g., multiple scalar fields realization \cite{Griest2002.PRD.66.123501}, everpresent $\Lambda$ \cite{Ahmed2004.PRD.69.103523}, and parameterized equation of state (EOS) \cite{Feng2006.PLB.634.101,Xia2005.MPLA.20.2409,Xia2006.PRD.74.083521,Pace2012.MNRAS.422.1186,Pan2018.PRD.98.063510}. Note that the Lagrangian can be reconstructed from a given EOS \cite{Nojiri2006.PLB.637.139,Zhao2006.PRD.73.123509,Zhao2007.ChinP.16.2830,SaezGomez2009.GravCos.15.134} and the result generally has a specified dependence on the Hubble constant $H_0$. This $H_0$-dependence is unattractive for a physical theory (see discussions in \cite{Frieman2008.ARAA.46.385,Peebles1988.ApJL.325.L17,Ratra1988.PRD.37.3406,Dodelson2000.PRL.85.5276}), and should not appear in the desired model. In \cite{Tian2020.PRD.101.063531}, we proposed a new oscillating dark energy model with the potential
\begin{equation}\label{eq:101}
  V(\phi)=V_0\exp\left[-\frac{\lambda_1+\lambda_2}{2}\phi-\frac{\alpha(\lambda_1-\lambda_2)}{2}\sin\frac{\phi}{\alpha}\right].
\end{equation}
Meanwhile we showed that the parameter space of $\{\lambda_1+\lambda_2>4$, $0<\lambda_2<0.39$, $\alpha=\mathcal{O}(1)$ and $V_0$ is arbitrary$\}$ is able to explain the late-time cosmic acceleration and solve the coincidence problem. In this model, the theoretically preferred Universe evolves in an oscillating scaling manner in the radiation epoch and in a chaotic accelerating manner in the matter epoch \cite{Tian2020.PRD.102.063509}. Mathematically, the radiation-matter transition induces a process of period-doubling bifurcation to chaos.

Besides the late-time acceleration, inflation is the other widely accepted accelerating phase in the Universe \cite{Guth1981.PRD.23.347}. These two accelerating phases may be driven by the same field. This idea emerged in the late 1980s \cite{Peebles1988.ApJL.325.L17,Ratra1988.PRD.37.3406} and a concrete model, named quintessential inflation, is proposed in \cite{Peebles1999.PRD.59.063505} after the observational confirmation of dark energy \cite{Riess1998.AJ.116.1009,Perlmutter1999.ApJ.517.565}. The MAU can also provide such a unified picture as its name implies. On this topic, previous work, e.g., \cite{Feng2006.PLB.634.101,Nojiri2006.PLB.637.139,SaezGomez2009.GravCos.15.134}, has studied the cosmic background evolution based on the parameterized oscillating EOS or its reconstructed Lagrangian. However, in the era of precision cosmology \cite{Akrami2020.AA.641.A10}, the performance of the MAU in inflation merits further discussion. In this paper, we study in detail the inflation model driven by the scalar field described by Eq. (\ref{eq:101}), and explore the general predictions of this kind of model.

This paper is organized as follows. In Sec. \ref{sec:02}, using the slow-roll approximation, we analyze the background evolution and the primordial inhomogeneities generated during inflation for our model. In particular, we prove an upper limit of the tensor-to-scalar ratio for a class of general MAU model. Section \ref{sec:03} analyzes the post-inflationary background evolution without considering the reheating process. Section \ref{sec:04} shows that all matters of the familiar Universe can be from gravitational reheating in our model. Discussion is presented in Sec. \ref{sec:05}. Throughout this paper, we adopt the SI units and inherit the notations in \cite{Tian2020.PRD.101.063531}, e.g., the slope $\lambda\equiv-V'/V$, $'\equiv\dx/\dx\phi$, $\phi$ is dimensionless and $[V]={\rm length}^{-2}$.

\section{Slow-roll inflation}\label{sec:02}
The early Universe is assumed to be dominated by the scalar field with the potential Eq. (\ref{eq:101}). As shown in Fig. \ref{fig:02}(a), the potential changes periodically between flat ($\lambda\approx\lambda_2$) and steep ($\lambda\approx\lambda_1$). A natural idea is that the flat part can drive the slow-roll inflation \cite{Linde1982.PLB.108.389,Albrecht1982.PRL.48.1220,Linde1983.PLB.129.177} while the steep part can end it. To be consistent with the slow-roll conditions but without loss of generality, we assume $\phi\approx\alpha\pi$ ($\lambda\approx\lambda_2$) at the beginning of inflation. With this setting, inflation would end around $\phi\approx2\alpha\pi$ ($\lambda\approx\lambda_1$). Here we present detailed analysis of this inflation process.

\subsection{Inflationary expansion}
The first issue we concern is the total e-folding number during inflation. The action of this physical system reads $S=\kappa^{-1}\int\dx^4x\sqrt{-g}\left[R/2-\p^\mu\phi\p_\mu\phi/2-V(\phi)\right]$, where $V(\phi)$ is given by Eq. (\ref{eq:101}). We further assume the Universe is flat, and then the cosmic evolution equations corresponding to the above action can be written as
\begin{subequations}\label{eq:201}
\begin{gather}
  H^2=\dot{\phi}^2/6+c^2V/3,\\
  \ddot{a}/a=-\dot{\phi}^2/3+c^2V/3,\\
  \ddot{\phi}+3H\dot{\phi}+c^2V'=0.
\end{gather}
\end{subequations}
Under the slow-roll conditions $\{\dot{\phi}^2\ll c^2V$ and $|\ddot{\phi}|\ll|c^2V'|\}$, Eq. (\ref{eq:201}) can be simplified to $H^2=c^2V/3$ and $3H\dot{\phi}=-c^2V'$. Dividing these two equations, we obtain
\begin{equation}\label{eq:202}
  \frac{1}{a}\frac{\dx a}{\dx\phi}=\frac{1}{\lambda}=\frac{2}{\lambda_1+\lambda_2+(\lambda_1-\lambda_2)\cos\frac{\phi}{\alpha}}.
\end{equation}
Integrating Eq. (\ref{eq:202}) gives the total e-folding number
\begin{equation}\label{eq:203}
  N_{\rm tot}=\int_{\phi_i}^{\phi_e}\frac{\dx\phi}{\lambda}
  =\frac{2\alpha}{\sqrt{\lambda_1\lambda_2}}\arctan(\sqrt{\frac{\lambda_2}{\lambda_1}}\left.\tan\frac{\phi}{2\alpha})\right|_{\phi_i}^{\phi_e},
\end{equation}
where the subscripts $i$ and $e$ denote the beginning and end of inflation respectively, and the second equality holds for $\phi_i\geqslant\alpha\pi$ due to the singularity of the tangent function. Note that the above result is insensitive to the value of $\phi_e$. For example, if $\lambda_2\ll1$, the integral from $3\alpha\pi/2$ to $2\alpha\pi$ is approximately equal to $2\alpha/\lambda_1$, which is much smaller than the leading term. For the case of $\phi_i=\alpha\pi$ and $\lambda_2\ll1$, Eq. (\ref{eq:203}) gives
\begin{equation}\label{eq:204}
  N_{\rm tot}=\frac{\alpha\pi}{\sqrt{\lambda_1\lambda_2}}+\mathcal{O}(1),
\end{equation}
where $\mathcal{O}(1)$ represents the minor influence of the exact value of $\phi_e$. Changing $\phi_i$ slightly does not affect the order of magnitude of the above result. Considering the typical values of $\lambda_1$ and $\alpha$ (see Sec. \ref{sec:01}), Eq. (\ref{eq:204}) requires $\lambda_2\lesssim\mathcal{O}(10^{-4})$ for successful inflation.

\subsection{Linear perturbations}
The second issue we concern is the primordial inhomogeneities generated during inflation. The source is quantum fluctuations \cite{Mukhanov1981.JETP.33.532,Hawking1982.PLB.115.295,Linde1982.PLB.116.335,Starobinsky1982.PLB.117.175,Guth1982.PRL.49.1110,Bardeen1983.PRD.28.679} and its evolution is governed by the Mukhanov-Sasaki equation \cite{Mukhanov1985.JETP.41.493,Sasaki1986.ProgTheorPhys.76.1036,Mukhanov1988.SovPhysJETP.67.1297}. The main result is characterized by spectral index and amplitude. The slow-roll approximation is suitable for dealing with this problem \cite{Liddle1992.PLB.291.391,Copeland1993.PRD.48.2529,Liddle1994.PRD.50.7222,Lidsey1997.RMP.69.373}. In our conventions, the potential slow-roll parameters are defined as $\epsilon_V\equiv1/2(V'/V)^2$ and $\eta_V\equiv V''/V$. We are concentrate on the lowest order result, where the spectral index of scalar perturbations $n_{\rm s}=1+2\eta_V-6\epsilon_V$ and the tensor-to-scalar ratio $r=16\epsilon_V$. The use of potential slow-roll parameters facilitates the analysis based on the potential Eq. (\ref{eq:101}).

\begin{figure}[!t]
  \centering
  \includegraphics[width=0.95\linewidth]{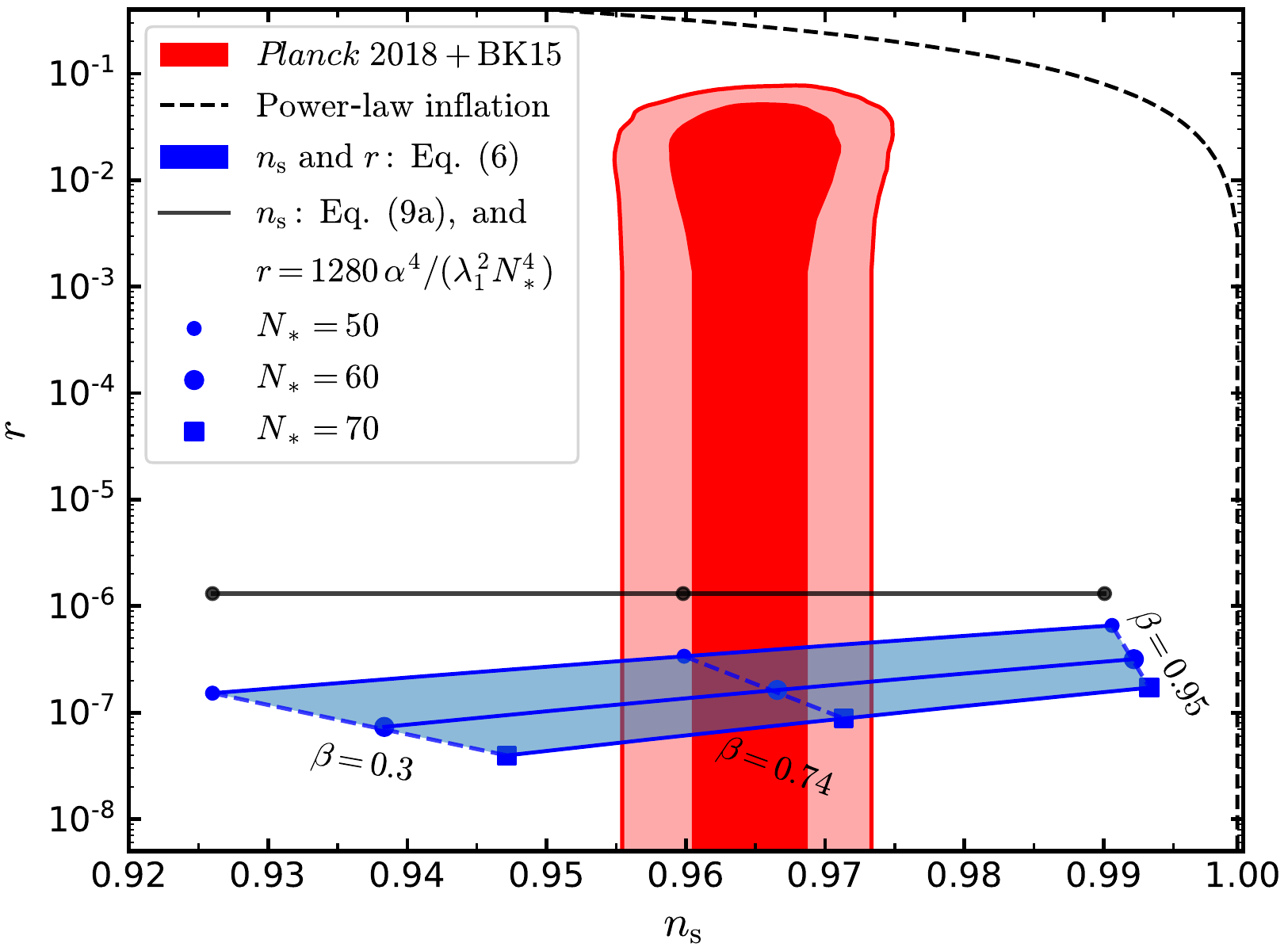}
  \caption{Predictions of the inflation model described by Eq. (\ref{eq:101}) in the plane $(n_{\rm s},r)$ together with \textit{Planck} 2018+BK15 measurements \cite{Akrami2020.AA.641.A10}. The red regions are the \textit{Planck} 2018+BK15 $68\%$ and $95\%$ CL contours. The blue region shows the result given by Eq. (\ref{eq:205}) with $\lambda_1=4.5$, $\alpha=0.6$, $50\leqslant N_\ast\leqslant70$ and $0.3\leqslant\beta\leqslant0.95$, which corresponds to $2.8\times10^{-5}\leqslant\lambda_2\leqslant2.9\times10^{-4}$ for $N_\ast=50$. We vary $\lambda_2$ in the calculations and mark the corresponding $\beta$ in the plot. The black solid line shows the result given by Eq. (\ref{eq:208}) with the same parameters and $\beta$-marks of the top blue line. For clarity, we multiply Eq. (\ref{eq:208b}) by $10$ to separate the black and blue plots. As Eq. (\ref{eq:101}) can be regarded as a modification of the exponential potential, we plot the predictions of the corresponding power-law inflation \cite{Martin2014.PDU.5.75} for comparison ($n_{\rm s}=1-r/8$, black dashed line).}
  \label{fig:01}
\end{figure}

At the level of approximation we are considering, for Eq. (\ref{eq:101}), direct calculation gives
\begin{subequations}\label{eq:205}
\begin{align}
  n_{\rm s}&=1-\left[\frac{(\lambda_1-\lambda_2)^2}{4}\cos^2\frac{\phi_\ast}{\alpha}+\frac{\lambda_1^2-\lambda_2^2}{2}\cos\frac{\phi_\ast}{\alpha}\right.\nonumber\\
    &\qquad\quad\left. -\frac{\lambda_1-\lambda_2}{\alpha}\sin\frac{\phi_\ast}{\alpha}+\frac{(\lambda_1+\lambda_2)^2}{4}\right], \label{eq:205a}\\
  r&=2\left[\lambda_1+\lambda_2+(\lambda_1-\lambda_2)\cos(\phi_\ast/\alpha)\right]^2,
\end{align}
\end{subequations}
where the subscript $\ast$ denotes horizon crossing. Integrating Eq. (\ref{eq:202}) from $\phi_\ast$ to $\phi_e$ gives
\begin{equation}\label{eq:206}
  N_\ast=-\frac{2\alpha}{\sqrt{\lambda_1\lambda_2}}\arctan(\sqrt{\frac{\lambda_2}{\lambda_1}}\tan\frac{\phi_\ast}{2\alpha}),
\end{equation}
in which we neglect the minor influence of the exact value of $\phi_e$ and assume $\phi_\ast$ is slightly larger than $\alpha\pi$. As we will see, this assumption is suitable for our analysis. Solving the above equation for $\phi_\ast$, we obtain
\begin{equation}\label{eq:207}
  \phi_\ast=2\alpha\left[\pi-\arctan(\sqrt{\lambda_1/\lambda_2}\tan\frac{\sqrt{\lambda_1\lambda_2}N_\ast}{2\alpha})\right].
\end{equation}
The range of $\phi_\ast$ given by Eq. (\ref{eq:207}) satisfies our requirement. In principle, for a given $N_\ast$, we can use Eq. (\ref{eq:207}) to calculate $\phi_\ast$, and then use Eq. (\ref{eq:205}) to calculate $n_{\rm s}$ and $r$. Note that $N_\ast$ given by Eq. (\ref{eq:206}) is the e-folding number from horizon crossing to inflation ending. For most of the inflation models, e.g., chaotic inflation \cite{Linde1983.PLB.129.177}, we have $50\lesssim N_\ast\lesssim60$ \cite{Akrami2020.AA.641.A10}. However, models that unifying inflation with dark energy may require a slightly larger $N_\ast$. For example, the quintessential inflation generally requires $N_\ast>60$ \cite{Dimopoulos2002.AstropartPhys.18.287,deHaro2019.JCAP.06.056}. In this section (especially in Fig. \ref{fig:01}), we assume $50\lesssim N_\ast\lesssim70$ for our model. The theoretical uncertainty of $N_\ast$ mainly comes from the modeling of the reheating process \cite{Martin2010.PRD.82.023511,Martin2014.PDU.5.75,Martin2021.arXiv.2105.03301}. In Sec. \ref{sec:04}, we present a specific reheating mechanism for our model. After that, a self-consistent calculation of $N_\ast$ is presented in the Appendix.

Before performing the calculations based on Eqs. (\ref{eq:205}) and (\ref{eq:207}), here we first use Taylor expansion to estimate the values of $n_{\rm s}$ and $r$. Substituting Eq. (\ref{eq:207}) into Eq. (\ref{eq:205a}), expanding the result around $\lambda_2=0$ and $N_\ast=+\infty$, and keeping the $\mathcal{O}(N_\ast^{-1})$ terms, we obtain
\begin{subequations}\label{eq:208}
\begin{equation}\label{eq:208a}
  n_{\rm s}=1-\frac{1}{N_\ast}\left(4-\frac{\pi^2\beta^2}{3}-\frac{\pi^4\beta^4}{180}-\frac{\pi^6\beta^6}{7560}+\cdots\right),
\end{equation}
where $\beta=\sqrt{\lambda_1\lambda_2}N_\ast/(\alpha\pi)$. Note that, inspired by Eq. (\ref{eq:204}), we assumed $\beta=\mathcal{O}(1)$ in the third step to derive Eq. (\ref{eq:208a}). \textit{Planck} 2018 observation gives $n_{\rm s}\approx0.965$ \cite{Akrami2020.AA.641.A10}, which corresponds to the expression in above parentheses approximately equal to $2$, i.e., $\beta\approx0.74$. Here $\beta<1$ ($N_\ast<\alpha\pi/\sqrt{\lambda_1\lambda_2}$) indicates $\phi_\ast>\alpha\pi$. Therefore, it is reasonable to assume $\beta=\mathcal{O}(1)$ and $\phi_\ast>\alpha\pi$ in the previous derivations. Considering the typical values of $\lambda_1$, $\alpha$ and $N_\ast$, $\beta\approx0.74$ requires $\lambda_2=\mathcal{O}(10^{-4})$. This result is consistent with the constraint given by Eq. (\ref{eq:204}) for successful inflation. For $r$, similar calculation gives
\begin{equation}\label{eq:208b}
  r=\mathcal{O}(1)\times\frac{128\alpha^4}{\lambda_1^2N_\ast^4},
\end{equation}
\end{subequations}
which indicates generally $r\sim10^{-7}$. \textit{Planck} 2018 \cite{Akrami2020.AA.641.A10} together with BICEP2/Keck Array BK15 \cite{Ade2018.PRL.121.221301} data requires $r<0.06$ at $95\%$ CL. Therefore, for both $n_{\rm s}$ and $r$, the predictions of the inflation model described by Eq. (\ref{eq:101}) and current observations are in good agreement.

Fig. \ref{fig:01} plots \textit{Planck} 2018+BK15 measurements and the results given by Eqs. (\ref{eq:205}) and (\ref{eq:208}). Comparing the ($n_{\rm s},r$) values of the three marked points in the blue and black solid lines with $N_\ast=50$, we conclude that Eq. (\ref{eq:208}) is a good approximation of Eq. (\ref{eq:205}). The red and blue regions confirm the agreement between theory and observations. Note that \textit{Planck} 2018+BK15 measurements rely on the $\Lambda$CDM model. For self-consistency, we should replace the $\Lambda$CDM model with the oscillating dark energy model in the parameter constraints. However, as shown in \cite{Pace2012.MNRAS.422.1186}, the influences of oscillating dark energy on the late-time evolution of cosmic inhomogeneities are generally small. For a rough estimate, it seems reasonable to adopt the $\Lambda$CDM-dependent results of the primordial inhomogeneities. The self-consistent constraints will be presented in a later paper.

The amplitude of the primordial inhomogeneities is related to the energy scale of inflation. In slow-roll approximation, the spectrum of scalar perturbations generated during inflation can be written as $\mathcal{P}_\mathcal{R}(k)=A_{\rm s}(k/k_\ast)^{n_{\rm s}-1}$, where the amplitude
\begin{equation}\label{eq:209}
  A_{\rm s}=\frac{\hbar\kappa H^2}{8\pi^2c\epsilon}\left(\frac{k_\ast c}{a_eH}\right)^{n_{\rm s}-1},
\end{equation}
and $k$ is the comoving wavenumber, $k_\ast$ denotes the characteristic wavenumber that crossed the horizon at $a=a_\ast$, the values of $\epsilon$ and $n_{\rm s}$ are evaluated at horizon crossing. In principle, $H$ should be evaluated at inflation ending as indicated by the exact solution of power-law inflation. However, the decrease in $H$ during inflation is small (especially in our model) and we can ignore its variation. In our conventions, we have
\begin{equation}
  H^2\approx\frac{c^2V_0}{3}\exp\left[-\frac{\lambda_1+\lambda_2}{2}\alpha\pi\right]
  \sim\frac{c^2V_0}{10^3},
\end{equation}
where the last approximation adopts the typical values of the parameters. The relation between $k_\ast$ and $N_\ast$ is
\begin{equation}
  e^{-N_\ast}\equiv\frac{a_\ast}{a_e}\approx\frac{a_\ast H_\ast}{a_eH_e}=\frac{k_\ast c}{a_eH_e}.
\end{equation}
Substituting the above results and $n_{\rm s}\approx1-2/N_\ast$ into Eq. (\ref{eq:209}), we obtain
\begin{equation}
  A_{\rm s}\approx\frac{\hbar\kappa H^2}{8\pi^2c\epsilon}e^2
  \approx2\frac{\hbar GH^2}{c^5\epsilon}.
\end{equation}
\textit{Planck} 2018 data gives $A_{\rm s}\approx2\times10^{-9}$ \cite{Akrami2020.AA.641.A10}. At the time of horizon crossing, considering $\epsilon=r/16\sim10^{-8}$ in our model, we obtain the following equivalent results
\begin{subequations}
\begin{align}
  H&\sim3\times10^{-9}\,t_{\rm P}^{-1},\label{eq:213a}\\
  V_0&\sim10^{-14}\,l_{\rm P}^{-2},\\
  \rho_{\rm inf}&\sim10^{-18}\,\rho_{\rm P},\label{eq:213c}
\end{align}
\end{subequations}
where $t_{\rm P}=\sqrt{\hbar G/c^5}$ is the Planck time, $l_{\rm P}=\sqrt{\hbar G/c^3}$ is the Planck length, $\rho_{\rm inf}$ is the energy density of inflaton and $\rho_{\rm P}=c^7/(\hbar G^2)$ is the Planck energy density. The above result shows $\rho_{\rm inf}\ll\rho_{\rm P}$ in our model. Actually, $\rho_{\rm inf}\ll\rho_{\rm P}$ is a general requirement for all the slow-roll inflation models. In the early 1980s, people thought this is a serious problem of the inflation theory (see the discussions in the Abstract of \cite{Hawking1982.PLB.115.295,Starobinsky1982.PLB.117.175,Guth1982.PRL.49.1110}). However, this issue has gradually lost people's attention in the follow-up development. The mainstream view of current cosmology community seems to admit that $\rho_{\rm inf}\ll\rho_{\rm P}$ is not a problem. We disagree with that. In our model, a much smaller $r$ does exacerbate the problem, i.e., we need a much smaller $\rho_{\rm inf}$. However, the essence of this problem remains unchanged.

\subsection{Smallness of $r$ in the general MAU}
Equation (\ref{eq:101}) behaves well in the inflationary stage. Especially, it predicts an extremely small $r$. It is natural to ask whether the small $r$ is a general consequence of the MAU. We can generalize Eq. (\ref{eq:101}) to
\begin{equation}\label{eq:214}
  V(\phi)=V_0\exp\left[-\frac{\lambda_1+\lambda_2}{2}\phi-\frac{\alpha(\lambda_1-\lambda_2)}{2}f(\frac{\phi}{\alpha})\right],
\end{equation}
where $f(x)$ is periodic with period approximately $2\pi$ and $-1\leqslant\dx f/\dx x\leqslant1$. The model proposed in \cite{Dodelson2000.PRL.85.5276} corresponds to $f\propto\ln[1+A\sin(B\phi)]$. Without loss of generality, we can assume $0\leqslant\lambda_2<\lambda_1$ and $\alpha=\mathcal{O}(1)$, which is necessary to solve the coincidence problem \cite{Tian2020.PRD.101.063531}. The slope of this potential satisfies $\lambda_2\leqslant\lambda\leqslant\lambda_1$. It is natural to assume that $\lambda$ is increasing from horizon crossing to inflation ending. Then, integrating Eq. (\ref{eq:202}) gives
\begin{equation}\label{eq:215}
  N_\ast=\int_{\phi_\ast}^{\phi_e}\frac{\dx\phi}{\lambda(\phi)}
  \lesssim\int_{\phi_\ast}^{\phi_e}\frac{\dx\phi}{\lambda_\ast}
  \approx\frac{1}{\lambda_\ast},
\end{equation}
where $\lambda_\ast=\lambda(\phi_\ast)$ and the last approximation uses $\alpha=\mathcal{O}(1)$. Equation (\ref{eq:215}) gives $\lambda_\ast\lesssim1/N_\ast$. Then we obtain
\begin{equation}
  r=16\epsilon=8\lambda_\ast^2\lesssim8/N_\ast^2.
\end{equation}
For the typical value $N_\ast=60$, the above equation gives $r\lesssim0.002$. The upper limit may change slightly in a specific model due to the missing coefficient. Conservatively, we may conclude $r<0.01$ for the general MAU model. For the traditional inflation models, some of them predict $r>0.01$, e.g., chaotic inflation \cite{Linde1983.PLB.129.177} and other models discussed in \cite{Boyle2006.PRL.96.111301}, while some predict $r<0.01$, e.g., Starobinsky $R^2$ inflation \cite{Starobinsky1980.PLB.91.99} and Higgs inflation \cite{Bezrukov2008.PLB.659.703}. Observational constraints on $r$ cannot distinguish our model from other traditional models that predict $r<0.01$. However, our calculations provide a physical motivation for the small $r$: if one want to use the MAU to solve the coincidence problem and unify inflation with dark energy in the framework of Eq. (\ref{eq:214}), then $r<0.01$.

Currently, \textit{Planck} 2018+BK15 data gives the strongest limit $r<0.06$ at $95\%$ CL \cite{Akrami2020.AA.641.A10}. There are some ongoing and upcoming cosmic microwave background polarization experiments to detect primordial gravitational waves, such as BICEP/Keck \cite{Ade2018.PRL.121.221301}, SPTpol \cite{Sayre2020.PRD.101.122003}, AliCPT \cite{Li2019.NSR.6.145}, LiteBIRD \cite{Hazumi2019.JLTP.194.443}, CMB-S4 \cite{Abazajian2020.arXiv.2008.12619} and so on. Part of these experiments may reach an upper limit of $r<0.001$. Observational confirmation of $r>0.01$ will make observations disfavor the MAU scenario. In contrast, if future observations limit $r$ to less than $0.01$, then this can be regarded as a circumstantial evidence for the MAU.

\section{Post-inflationary evolution}\label{sec:03}
Generally, the single scalar field slow-roll inflation ends once the potential becomes steep, and then the scalar field oscillates near the minimum of its potential \cite{Bassett2006.RMP.78.537}. However, things will be different in our model as there is no local minimum of the potential (note that $0<\lambda_2\leqslant\lambda\leqslant\lambda_1$). A monotonic potential will drive the scalar field roll to infinity. Here we present a quantitative analysis of this process. In this section, reheating and its backreaction on the background dynamics \cite{Bassett2006.RMP.78.537} are not taken into account.

Here the slow-roll approximation is no longer applicable due to the non-negligible kinetic energy of the scalar field. Inspired by \cite{Tian2020.PRD.102.063509}, what we need may be the dynamical system form of Eq. (\ref{eq:201}). Introducing the dimensionless variables $x_1\equiv\dot{\phi}/(\sqrt{6}H)$ and $\nu\equiv\sqrt{6}(\lambda^2-V''/V)$, Eq. (\ref{eq:201}) can be rewritten as
\begin{subequations}\label{eq:301}
\begin{align}
  \frac{\dx x_1}{\dx N}&=(1-x_1^2)(\frac{\sqrt{6}}{2}\lambda-3x_1), \label{eq:301a}\\
  \frac{\dx\lambda}{\dx N}&=\nu x_1,\\
  \frac{\dx\nu}{\dx N}&=\frac{3x_1}{\alpha^2}(\lambda_1+\lambda_2-2\lambda),
\end{align}
\end{subequations}
where $N\equiv\ln(a/a_i)$. A constraint equation for $\lambda$ and $\nu$ is given by Eq. (9) in \cite{Tian2020.PRD.101.063531}. To characterize the evolution of this system, we use the EOS $w_\phi=2x_1^2-1$ \cite{Tian2020.PRD.101.063531} and the Hubble slow-roll parameters \cite{Liddle1994.PRD.50.7222}
\begin{subequations}
\begin{align}
  \epsilon_H&\equiv-\frac{\dot{H}}{H^2}=\frac{\dot{\phi}^2}{2H^2}=3x_1^2, \label{eq:302a}\\
  \eta_H&\equiv-\frac{\ddot{\phi}}{H\dot{\phi}}=3-\frac{\sqrt{6}\lambda(1-x_1^2)}{2 x_1},
\end{align}
\end{subequations}
where the above derivations following the definitions use Eq. (\ref{eq:201}). To lowest order, we have $\epsilon_H=\epsilon_V$, $\eta_H=\eta_V-\epsilon_V$ and $n_{\rm s}=1+2\eta_H-4\epsilon_H$. The use of Hubble slow-roll parameters facilitates the analysis based on the dynamical system Eq. (\ref{eq:301}).

\begin{figure}[!t]
  \centering
  \includegraphics[width=0.95\linewidth]{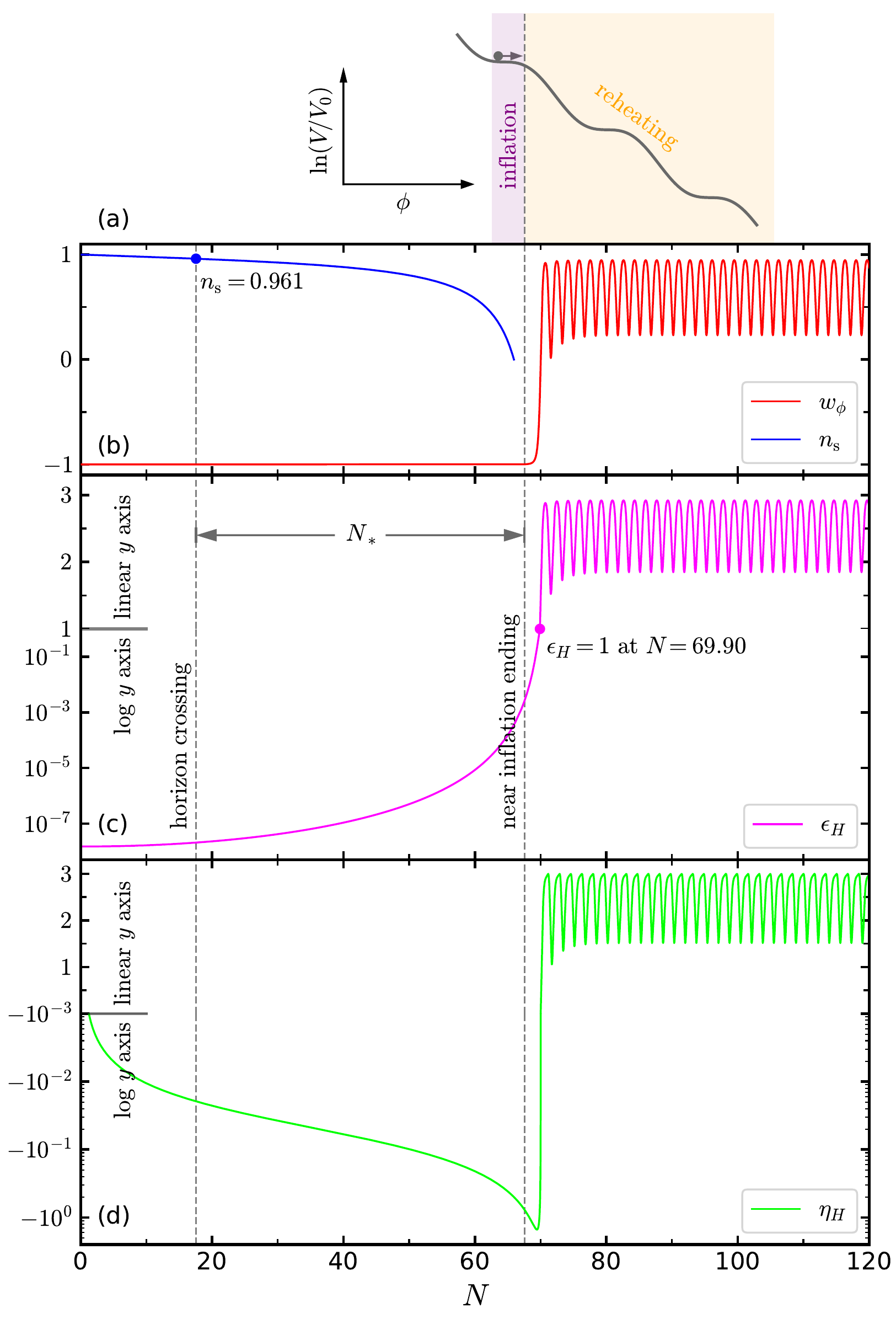}
  \caption{Evolution of $w_\phi$, $\epsilon_H$ and $\eta_H$. The model parameters are $\lambda_1=4.5$, $\lambda_2=1.73\times10^{-4}$ ($\beta=0.74$), $\alpha=0.6$ and $N_\ast=50$. For the initial conditions, we assume $\phi_i=\alpha\pi$, which corresponds to $\lambda=\lambda_2$ and $\nu=0$ at the beginning of inflation. In addition, we assume the slow-roll initial condition for $x_1$, and then Eq. (\ref{eq:302a}) gives $x_1=\sqrt{\epsilon_V(\phi_i)/3}=\lambda_2/\sqrt{6}$. The second vertical dashed line corresponds to $N=\alpha\pi/\sqrt{\lambda_1\lambda_2}=67.56$ [near inflation ending, see Eq. (\ref{eq:204})]. Strictly speaking, inflation ends at $N=69.90$, at which $\epsilon_H=1$ and $\ddot{a}/a=0$. The first vertical dashed line corresponds to $N=67.56-N_\ast$ (horizon crossing). In the subplot (d), the minor ticks from $-10^{-i}$ to $-10^{-i-1}$ correspond to $-10^{-i}\times(1/2,1/3,\cdots,1/9)$ respectively. The spectral index $n_{\rm s}=1+2\eta_H-4\epsilon_H$ is given in the subplot (b). For the above settings, $n_{\rm s}=0.961$ at horizon crossing [meanwhile Eq. (\ref{eq:208a}) gives a consistent result $n_{\rm s}=0.960$]. The motion of the scalar field in its potential is illustrated in the subplot (a).}
  \label{fig:02}
\end{figure}

To get a first glance of the system's property, based on Eq. (\ref{eq:301}), Fig. \ref{fig:02} plot the evolution of $w_\phi$, $\epsilon_H$ and $\eta_H$ for a representative set of model parameters. This figure shows that inflation ends naturally in our model as we expected. Especially, inflation will not happen again when the scalar field passes through the second flat part ($\phi\approx3\alpha\pi$ in our settings, see the first paragraph in Sec. \ref{sec:02}). The reason is the scalar field accumulates sufficient kinetic energy as it passes through the first steep part ($\phi\approx2\alpha\pi$) and then it will quickly pass through the following flat parts.

After the end of inflation, Fig. \ref{fig:02} shows the existence of oscillation. Here we investigate the oscillation frequency and the average EOS $\overline{w}_\phi$ during oscillation. Similar to \cite{Tian2020.PRD.101.063531,Tian2020.PRD.102.063509}, the result of exponential potential may provide direct clues. For $V(\phi)\propto\exp(-\lambda\phi)$, where $\lambda$ is a positive constant, the system's evolution equation is given by Eq. (\ref{eq:301a}), and the stable critical point is $x_1=\lambda/\sqrt{6}$ if $\lambda<\sqrt{6}$ and $x_1=1$ if $\lambda\geqslant\sqrt{6}$ \cite{Copeland1998.PRD.57.4686}. For the dynamical system Eq. (\ref{eq:301}), considering $w_\phi=2x_1^2-1$ and the results obtained for the pure exponential potential, we may be able to find an approximate EOS
\begin{equation}\label{eq:303}
  \overline{w}_{\phi,{\rm app}} = \left\{ \begin{array}{ll}
    (\lambda_1+\lambda_2)^2/12-1 & \textrm{if $\lambda_1+\lambda_2<2\sqrt{6}$}, \\
    1 & \textrm{if $\lambda_1+\lambda_2\geqslant2\sqrt{6}$},
\end{array} \right.
\end{equation}
during oscillation. As $w_\phi\leqslant1$ for all the quintessence models, we guess that the oscillation only occurs when $\overline{w}_{\phi}<1$, i.e., approximately $\lambda_1+\lambda_2<2\sqrt{6}$. Otherwise, if $\overline{w}_\phi=1$, there will be no upper room for $w_\phi$ to be oscillating. To find an approximation of the oscillation frequency within this parameter space, we can use the Fourier series method to solve Eq. (\ref{eq:301}). Details can be found in the appendix of \cite{Tian2020.PRD.102.063509}, and the only difference is that the background values should be $x_1=(\lambda_1+\lambda_2)/\sqrt{24}$, $\lambda=(\lambda_1+\lambda_2)/2$ and $\nu=0$. Note that this setting is consistent with the $\lambda$-$\nu$ constraint given by Eq. (9) in \cite{Tian2020.PRD.101.063531} as the \emph{background} means $\lambda_1=\lambda_2$. The result of the oscillation frequency is
\begin{equation}\label{eq:304}
  f_{\rm os}=\frac{\lambda_1+\lambda_2}{4\alpha\pi}.
\end{equation}
Note that, in principle, we can only prove Eqs. (\ref{eq:303}) and (\ref{eq:304}) hold for $\lambda_1-\lambda_2\ll1$, which is outside the model's viable parameter space.

\begin{figure*}[!t]
  \centering
  \includegraphics[width=0.49\linewidth]{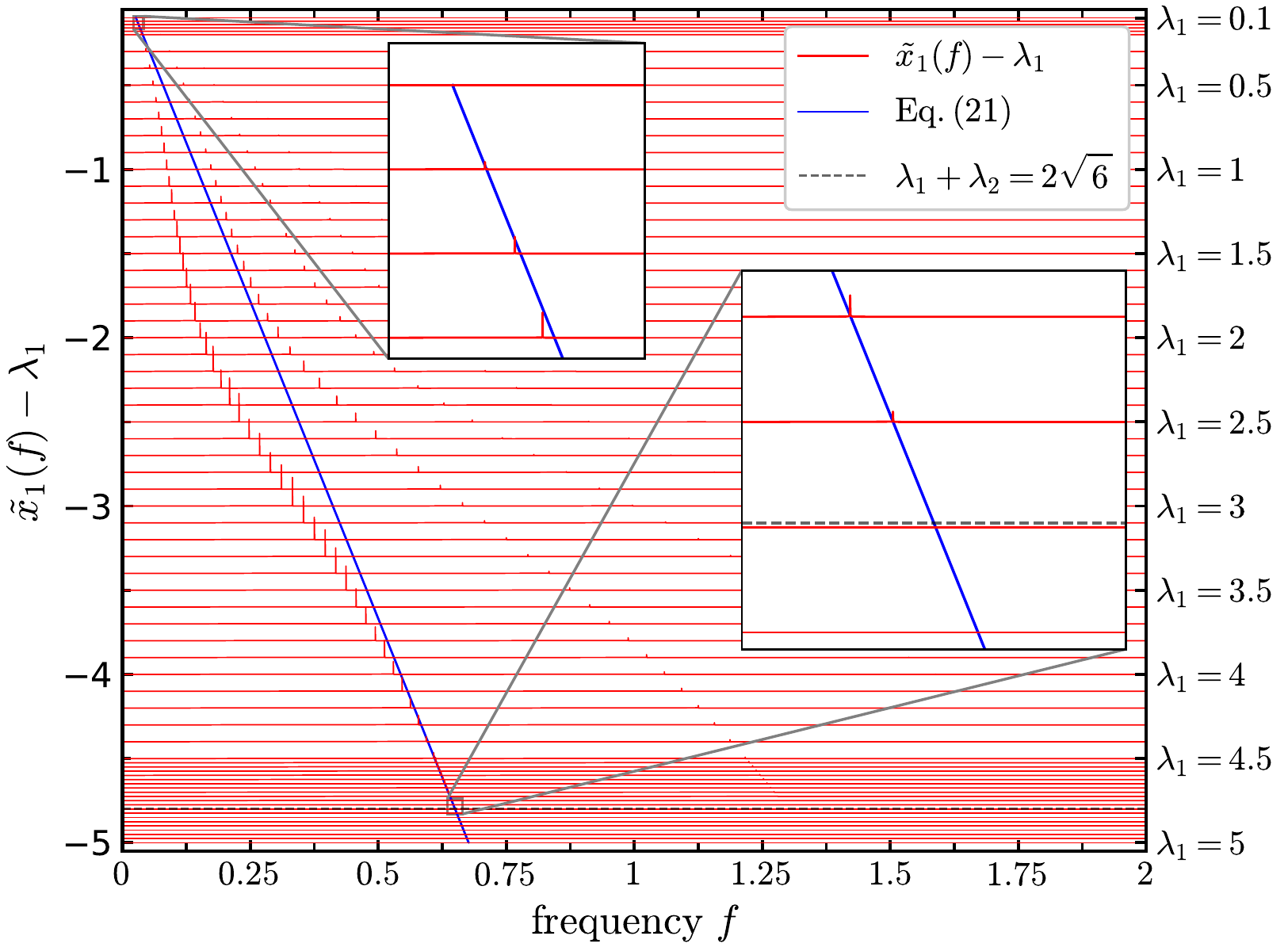}
  \includegraphics[width=0.49\linewidth]{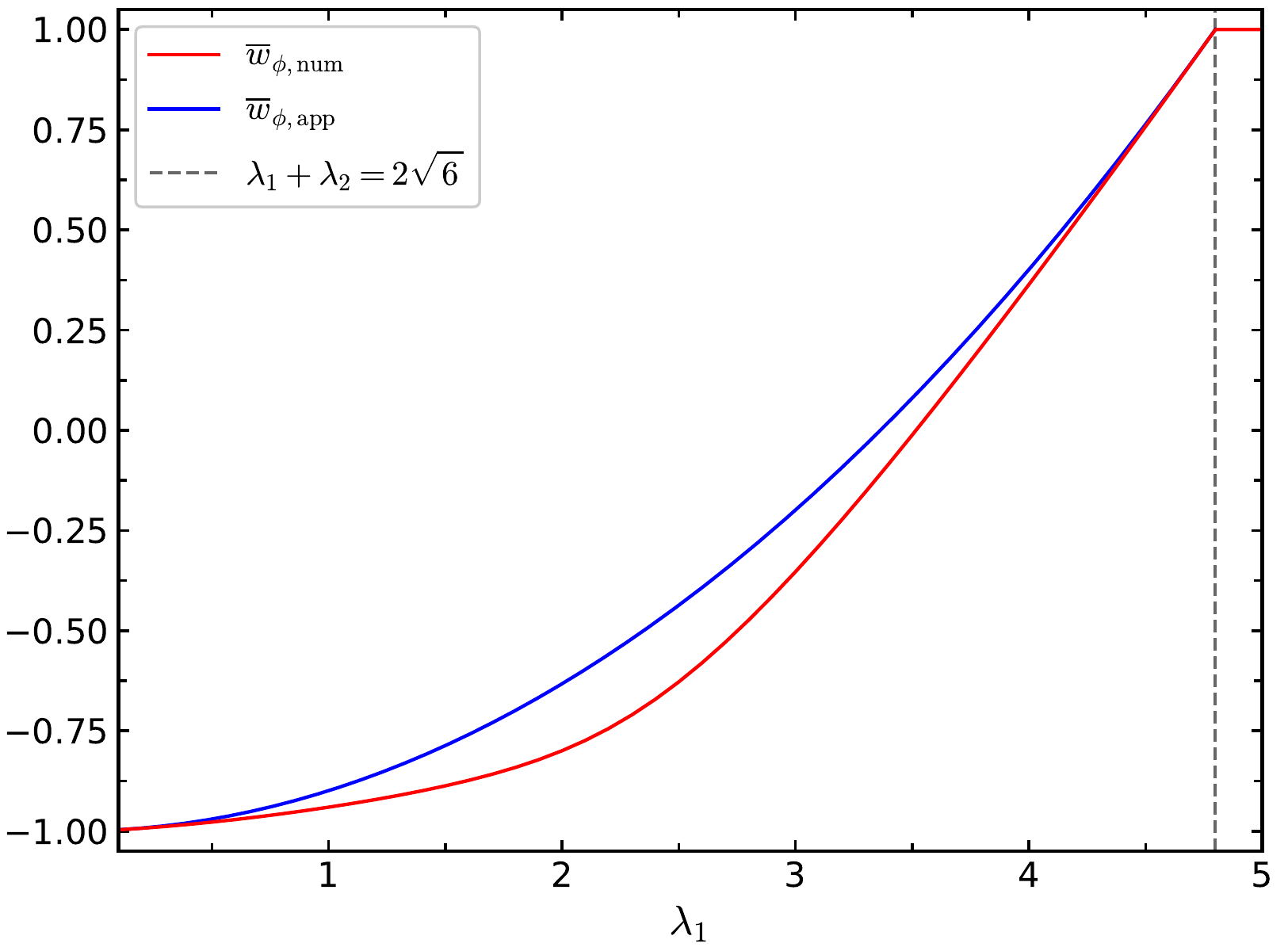}
  \caption{Left: Fourier transform of $x_1(N)$. The result is normalized by the signal length. What we concern is the location of the peak rather than its amplitude. Right: $\overline{w}_\phi$ versus $\lambda_1$. The numerical result is given by $\overline{w}_{\phi,{\rm num}}=(N_2-N_1)^{-1}\int_{N_1}^{N_2}w_\phi(\tilde{N})\dx\tilde{N}$, which characterizes the density decrease rate (see Eq. (1) in \cite{Tian2021.PRD.103.043518}).  The model parameters are $\lambda_2=0.1$, $\alpha=0.6$ and $\lambda_1$ can be found in the figures. The initial conditions are $x_1=0.5$, $\lambda=(\lambda_1+\lambda_2)/2$ and $\nu=\nu_+(\lambda)$ (see Eq. (9) in \cite{Tian2020.PRD.101.063531}). We numerically solve Eq. (\ref{eq:301}) in $N\in[0,10^5]$, perform Fourier transform and calculate $\overline{w}_{\phi,{\rm num}}$ in $N\in[500,10^5]$. The numerical results are plotted in red, while Eqs. (\ref{eq:303}) and (\ref{eq:304}) are plotted in blue. The gray dashed line corresponds to the boundary $\lambda_1+\lambda_2=2\sqrt{6}$.}
  \label{fig:03}
\end{figure*}

Fig. \ref{fig:03} plots the numerical results of $\overline{w}_\phi$ and the Fourier transform $\tilde{x}_1(f)$, together with Eqs. (\ref{eq:303}) and (\ref{eq:304}). When $\lambda_1\approx\lambda_2$ ($\lambda_1-\lambda_2\ll1$), for both $\overline{w}_\phi$ and $f_{\rm os}$, the numerical and approximate results are in good agreement as we expected. The difference between the numerical results and the approximations grows as $\lambda_1$ increases. However, the difference miraculously becomes negligible when $\lambda_1\gtrsim4$. Especially, the boundary $\lambda_1+\lambda_2=2\sqrt{6}\approx4.899$ is verified. Therefore, for the viable parameter space ($\lambda_1+\lambda_2>4$), Eqs. (\ref{eq:303}) and (\ref{eq:304}) are good approximations of the exact results. Equation (\ref{eq:303}) together with $\lambda_1+\lambda_2>4$ give $\overline{w}_\phi>1/3$, i.e., $\overline{w}_\phi$ is larger than the EOS of radiation when the Universe is dominated by the oscillating scalar field. If radiation is generated during this period, then its relative energy density will increase with cosmic expansion and the Universe will gradually enter the radiation era. Fig. \ref{fig:03} also shows that there is no period-doubling bifurcation and chaos in the three-dimensional dynamical system Eq. (\ref{eq:301}). This is different with the properties of the four-dimensional dynamical system discussed in \cite{Tian2020.PRD.102.063509}.

Terminology might be mentioned. In the literature, the decelerating post-inflationary period may be called ``deflationary" \cite{Spokoiny1993.PLB.315.40} or ``kination" \cite{Joyce1997.PRD.55.1875}. In our model, the post-inflationary dynamics is dominated by the scalar field with an average EOS $\overline{w}_\phi>1/3$. In this period, both the kinetic and potential energy of the scalar field are not negligible. Therefore, we call this period ``deflation" instead of ``kination".

\section{Gravitational reheating}\label{sec:04}
At the end of the inflation, the Universe is very cold due to the fast cosmic expansion. A mechanism is needed to reheat the Universe \cite{Guth1981.PRD.23.347}. In modern cosmology, the reheating is typically assumed to occur through the decay of inflaton, which is sourced by the nonminimal coupling of inflaton and other fields \cite{Dolgov1982.PLB.116.329,Abbott1982.PLB.117.29,Albrecht1982.PRL.48.1437,Kofman1994.PRL.73.3195,Kofman1997.PRD.56.3258,Peter2013.book}. A less commonly discussed case is gravitational reheating, i.e., particle creation in curved spacetime. This scenario was first investigated in \cite{Ford1987.PRD.35.2955} and applied to the quintessential inflation \cite{Peebles1999.PRD.59.063505}. One important feature of the gravitational reheating is that it does not need to introduce any nonminimal coupling. Here we study whether the gravitational reheating can successfully create the hot Universe in our model. Especially, we concentrate on the radiation temperature $T_{\rm reh}$ at the end of deflation, i.e., the beginning of radiation era. The requirement is $T_{\rm reh}\gtrsim10^{11}\,{\rm K}$ ($\sim10\,{\rm MeV}$ in natural units, i.e., the temperature at which primordial nucleosynthesis starts \cite{Martin2010.PRD.82.023511}).

The Universe is assumed to undergo a transition from inflation (nearly de Sitter spacetime) to deflation (decelerating phase). For a massless scalar field, the created energy density at the end of inflation from the inflation-deflation transition is
\begin{equation}\label{eq:401}
  \rho_{\rm r}(a_e)=R\rho_{\rm inf}^2/\rho_{\rm P},
\end{equation}
where $R\sim0.01$ for arbitrary power-law deflation \cite{Ford1987.PRD.35.2955,Giovannini1998.PRD.58.083504}, and $\rho_{\rm inf}$ is given by Eq. (\ref{eq:213c}) in our model. This creation can be regarded as an instantaneous process. Based on the calculations presented in Sec. \ref{sec:03}, it is natural to assume that Eq. (\ref{eq:401}) applies to our model. There may be many scalar fields in the Universe and each one contributes $0.01$ to $R$. Following \cite{Peebles1999.PRD.59.063505}, hereafter we assume $0.01\lesssim R\lesssim1$, which corresponds to $T_{\rm r}(a_e)\sim10^{23}\,{\rm K}$, where $T_{\rm r}(a_e)$ is the radiation temperature at the end of inflation (after reheating). During deflation, the radiation energy density $\rho_{\rm r}=\rho_{r}(a_e)\cdot(a_e/a)^4$ and the inflaton energy density $\rho_\phi\approx\rho_{\rm inf}\cdot(a_e/a)^{3(1+\overline{w}_\phi)}$, where $1/3<\overline{w}_\phi\leqslant1$. Thus, the ratio of $\rho_{\rm r}$ and $\rho_\phi$ is
\begin{equation}
  \rho_{\rm r}/\rho_\phi\approx10^{-18}\cdot R\cdot(a/a_e)^{3\overline{w}_\phi-1}.
\end{equation}
Radiation era thus begins at the scale factor
\begin{equation}
  a_{\rm reh}\approx a_e\cdot(10^{18}/R)^{1/(3\overline{w}_\phi-1)}.
\end{equation}
Meanwhile the radiation temperature $T_{\rm reh}$ is
\begin{align}
  T_{\rm reh}&=\left[\frac{15c^3\hbar^3\rho_{\rm r}(a_{\rm reh})}{\pi^2k_{\rm B}^4}\right]^{1/4}\nonumber\\
  &=\left\{ \begin{array}{ll}
    1.6\times10^{14}R^{0.75}\,{\rm K} & \textrm{if $\overline{w}_\phi=1$,}\\
    4.1\times10^{12}R^{0.84}\,{\rm K} & \textrm{if $\overline{w}_\phi=0.9$,}\\
    2.2\times10^{10}R^{0.96}\,{\rm K} & \textrm{if $\overline{w}_\phi=0.8$,}
  \end{array} \right.
\end{align}
where $k_{\rm B}$ is the Boltzmann constant. Therefore, it is possible to realize $T_{\rm reh}\gtrsim10^{11}\,{\rm K}$ with suitable parameters in our model. For example, parameters with $\overline{w}_\phi>0.825$ if $R=1$ and $\overline{w}_\phi>0.903$ if $R=0.01$. Considering Eq. (\ref{eq:303}) and $\lambda_2\ll1$, we obtain $\lambda_1>4.68$ and $\lambda_1>4.78$ respectively.

\section{Discussion}\label{sec:05}
Regarding the late-time acceleration as one of the accelerating phases in the MAU provides an attractive solution to the coincidence problem. Meanwhile, MAU provides a natural unification of inflation and dark energy. In this paper, we analyze the inflationary consequences of the MAU model described by Eq. (\ref{eq:101}). This model with $\lambda_2=\mathcal{O}(10^{-4})$ gives sufficient e-folding number, suitable $n_{\rm s}$ and extremely small $r$. The post-inflationary expansion is decelerating with an average EOS $\overline{w}_\phi>1/3$. In this framework, we show that gravitational particle creation at the end of inflation is sufficient to reheat the hot Universe.

Together with our previous works, here we summarize two observational probes for the general MAU scenario. One probe is the primordial gravitational waves: MAU generally predicts $r<0.01$. The physical origin of this upper limit is that the coincidence problem requires the increment of $\phi$ cannot be too large during one accelerating phase (including inflation) \cite{Tian2020.PRD.101.063531}. During inflation, in order to obtain sufficient e-folding number, the potential must be very flat, which results in small $\epsilon_V$ and $r$. The other probe is the Hubble expansion rate at cosmic dawn (denoted as $H_{\rm cd}$) measured by the 21\,cm signals \cite{Tian2020.PRD.102.063509,Munoz2019.PRL.123.131301}. In the MAU, the scalar field is comparable with the normal matters across the whole cosmic history, and thus $H_{\rm cd}$ should be much larger than that in the $\Lambda$CDM model. We expect future observations can prove or disprove these two predictions. But note that the MAU is not the unique model to obtain small $r$ or large $H_{\rm cd}$. For example, Starobinsky $R^2$ inflation also gives small $r$ \cite{Starobinsky1980.PLB.91.99,Akrami2020.AA.641.A10} and some interacting dark energy models may give large $H_{\rm cd}$ \cite{Li2020.PLB.801.135141}. What we have done is to provide a physical motivation (i.e., the cosmological coincidence problem) for the small $r$ and large $H_{\rm cd}$.

There are several unnatural features of our model. In principle, arbitrary $V_0$ can be used to explain the late-time cosmic acceleration \cite{Tian2020.PRD.101.063531}. However, in order to obtain a suitable $A_{\rm s}$, we have to specify an extremely small value for $V_0$ relative to the Planck scale values. This is unnatural (a fine-tuning problem), as in the conventional inflation models \cite{Hawking1982.PLB.115.295,Starobinsky1982.PLB.117.175,Guth1982.PRL.49.1110}. In addition, the dimensionless parameter $\lambda_2=\mathcal{O}(10^{-4})$ indicates the other minor fine-tuning problem in our model. We believe that nature (cosmology) prefers models that only introduce Planck scale parameters and dimensionless parameters of order unity. Inspired by \cite{Peebles1999.PRD.59.063505}, those two fine-tuning problems may can be addressed in models with an extra inflation field $\chi$. In the new model, both $V_0$ and $\lambda_2$ are not constants but slowly varying functions of $\chi$. This possibility will be explored in the future.

\section*{Acknowledgements}
This work was supported by the National Natural Science Foundation of China under Grants Nos. 11633001, 11920101003 and 12021003, the Strategic Priority Research Program of the Chinese Academy of Sciences under Grant No. XDB23000000 and the Interdiscipline Research Funds of Beijing Normal University. S.X.T. was supported by the Initiative Postdocs Supporting Program under Grant No. BX20200065.

\appendix
\section{Self-consistent calculation of $N_\ast$}\label{sec:Appendix}
In Sec. \ref{sec:02}, we assumed $50\leqslant N_\ast\leqslant70$ without considering the post-inflationary dynamics. However, specification of the reheating mechanism (see Sec. \ref{sec:04}) enables us to self-consistently calculate $N_\ast$. The purpose of this Appendix is to do so. As we show in the following, our model requires a slightly larger $N_\ast$ ($N_\ast\gtrsim60$), which is similar to the result obtained for the quintessential inflation \cite{Dimopoulos2002.AstropartPhys.18.287,deHaro2019.JCAP.06.056}. This similarity mainly comes from the same reheating mechanism and similar deflationary dynamics in our model and the quintessential inflation.

During inflation, the mode $k_\ast$ crossed the horizon at $k_\ast c=a_\ast H_\ast$. The standard way to calculate $N_\ast$ is to write down
\begin{align}\label{eq:A01}
  \frac{k_\ast c}{a_0H_0}=\frac{a_\ast H_\ast}{a_0H_0}
  =\frac{a_\ast}{a_e}\frac{a_e}{a_0}\frac{H_\ast}{H_0},
\end{align}
where the subscript $0$ means today. Compared with Eq. (5.13) in \cite{Liddle1993.PhysRep.231.1}, here we do not introduce the scale factor $a_{\rm reh}$. The reason is, as we mentioned in Sec. \ref{sec:04}, the reheating is instantaneous at the end of inflation. The created radiation energy density is given by Eq. (\ref{eq:401}). After reheating, the radiation energy density is proportional to $a^{-4}$, which gives
\begin{equation}
  \frac{a_e}{a_0}=\left[\frac{\rho_{\rm r}(a_0)}{\rho_{\rm r}(a_e)}\right]^{1/4}
  =\exp\left(-52.4-\frac{1}{4}\ln R\right).
\end{equation}
Equation (\ref{eq:213a}) gives $H_\ast/H_0=e^{120.6}$, where we adopt $H_0=70\,{\rm km}/{\rm s}/{\rm Mpc}$. Substituting the above results into Eq. (\ref{eq:A01}), we obtain
\begin{equation}
  N_\ast\equiv\ln\frac{a_e}{a_\ast}
  =68.2-\frac{1}{4}\ln R-\ln\frac{k_\ast c}{a_0H_0}.
\end{equation}
For the typical values $0.01\leqslant R\leqslant1$ and $k_\ast/a_0=0.002\,{\rm Mpc}^{-1}$ \cite{Akrami2020.AA.641.A10}, we obtain $67.2\geqslant N_\ast\geqslant66.1$.

\end{document}